\documentclass[showpacs,preprintnumbers,amsmath,amssymb]{revtex4}

\usepackage{graphicx}
\usepackage{dcolumn}
\usepackage{bm}
\usepackage{epsfig}

\newcommand{\dhd}{{\textstyle d}
\lower.03ex\hbox{\kern-0.40em$^{\scriptstyle-}$}\kern-0.08em{}}

\newcommand{\calo}{{\cal O}}

\newcommand{\calz}{{\cal Z}}

\newcommand{\halo}{\hat{\cal O}} 

\begin{document}

~~~~~~~~~~~~~~~~~~~~~~~~~~~~~~~~~~~~~~~~~~~~~~~~~~~~~~~~~
\preprint{CPHT-006.0110; LPT-ORSAY 10-05}

\title{Small-x Evolution in the Next-to-Leading Order\footnote{Talk given at the workshop 
"Recent Advances in Perturbative QCD and Hadronic Physics" Trento, 20-25 July 2009:
honoring the 75th birthday of Prof. A. V. Efremov.}
}

\author{ Giovanni Antonio Chirilli}
\affiliation{Physics Department, Old Dominion University,
Norfolk, VA 23529, and\\
Theory Group, JLAB, 12000 Jefferson Ave, Newport News, VA 23606, USA\\
CPHT Ecole Polytechnique, 91128 Palaiseau cedex, France\\
Bâtiment 210 Univ. Paris-Sud 11, 91405 Orsay Cedex, France\\
E-mail: chirilli@jlab.org; chirilli@cpht.polytechnique.fr
}

\begin{abstract}
After a brief introduction to Deep Inelastic Scattering in the Bjorken limit and in the Regge Limit we discuss the
operator product expansion in terms of non local string operator and in terms of Wilson lines.
We will show how the high-energy behavior of amplitudes in gauge theories can be reformulated in terms of the evolution of Wilson-line
operators. In the leading order this evolution is governed by the non-linear Balitsky-
Kovchegov (BK) equation. In order to see if this equation is relevant for existing or
future deep inelastic scattering (DIS) accelerators (like Electron Ion Collider (EIC) or
Large Hadron electron Collider (LHeC)) one needs to know the next-to-leading 
order (NLO) corrections. In addition, the NLO corrections define the
scale of the running-coupling constant in the BK equation and therefore determine
the magnitude of the leading-order cross sections. 
In Quantum Chromodynamics (QCD),
the next-to-leading order BK equation has both conformal and non-conformal parts.
The NLO kernel for the composite operators
resolves in a sum of the conformal part and the running-coupling part.
The QCD and ${\cal N}=4$ SYM kernel of the BK equation is presented. 
\keywords{Small-x evolution; Wilson line; High-density QCD.}
\end{abstract}

\pacs{12.38.Bx, 12.38.Cy}

\maketitle

\section{Introduction} 
It is well established that hadrons are made of asymptotically free particles which are quasi on-shell excitations.
Their confining properties does not allow to measure them directly, and therefore Deep Inelastic Scattering (DIS)
experiments are performed to study
the structure functions of hadrons related to lowest order in QCD to the number density constituents.
In DIS a lepton with momentum $k$ scatter off a parton inside a hadron with momentum $P$
through an exchange of a virtual photon with virtuality $Q^2=-q^2$ which probes the structure 
of the target. From the measurements of the final momentum $k'$ of the scattered lepton one obtains the momentum $q=k'-k$ 
transferred by the virtual photon to the hadron. 
A suitable frame in which the parton picture is revealed is the infinite-momentum frame of
the hadron. In this frame we observe different time scale for the interaction of the projectile with the constituents of the target and the 
interaction among the constituents of the target. Furthermore, Lorentz contraction in the longitudinal direction reduces the target into a "pancake"
and time dilation enhances the lifetime of the fluctuating partons so that the
hadron constituents are effectively "frozen" during the scattering with the projectile. In the limit in which the effective mass of the partons is negligible
and therefore have small rest mass (partons are asymptotically free) the interactions projectile-partons can be considered incoherent.
In the Breit frame the longitudinal momentum of the virtual photon is zero ($q^\mu=(q_0, q_\perp, 0)$) and
$Q^2=q^2_\perp$. In this case the virtual photon acts as a microscope resolving partons in the transverse plane of dimension
of the order of ${1\over q^2_\perp}$.
In the limit in which the energy and the momentum transfer to the target go
to infinity while their ratio is kept fixed (the Bjorken limit), the structure functions evolve according to the DGLAP 
(Dokshitzer, Gribov, Lipatov, Altarelli, Parisi) equation\cite{dglap}
which is an evolution equation in $Q^2$:
the power resolution of the virtual photon probing the target is inversely proportional
to the momentum transfer; the higher is the momentum transfer the smaller is the size of the parton 
resolved by the virtual photon in the transverse direction. 
The DGLAP equation, obtained by resumming radiative corrections proportional to $(\alpha_s \ln Q^2)^n$ for all $n\geq 1$, 
predicts the increase of the parton distributions and describe the evolution of the system
towards a dilute regime. Besides, although such evolution equation has been successful 
in explaining many features of HERA data like the increase of the parton distribution, 
in the region of low momentum transfer its predictive power is lost and consequently higher twist corrections ($1/Q^2$ corrections) 
are not negligible anymore.
It has been observed in experiments like H1 or Zeus at Hera 
that the structure functions of gluons and sea quarks 
grow with decreasing values of the Lorentz invariant variable $x_B$\cite{rapid-grow}.
The small-$x_B$ region is obtained in the so called Regge limit.
When the center of the mass energy square $s$ 
and the momentum transfer $Q^2$ are much greater then the mass of the hadron,
the Bjorken variable is approximately given by $x_B\equiv {Q^2\over 2P\cdot q}\sim{Q^2\over s}$. The Regge limit is then achieved when
the parameter $s$ is taken very large while keeping the momentum transfer $Q^2$ fixed, consequently $x_B$ decreases. 
At high energies (Regge limit) contributions proportional to $(\alpha_s \ln s)^n$ for all $n\geq 1$ cannot 
be neglected anymore, and therefore they have to be resummed. The equation governing the evolution of the gluon distribution, and
of the scattering amplitude at high energies is the BFKL\cite{bfkl} (Balitsky, Fadin, Kuraev, Lipatov) equation.
As previously mentioned, the DGLAP equation is an evolution equation towards a dilute regime, 
and it needs higher twist corrections for low values of $Q^2$. 
These type of corrections have some theoretical difficulties. On the other hand, 
if we fix the momentum transfer and we let the hadron evolve with rapidity, the
parton density increases. Fixing the momentum $Q^2$ means that the virtual photon resolves in the hadron
partons with fixed size in the transverse direction while their number increases. The BFKL equation provides a nice 
description of such an increase in density, but it fails to describe the evolution of parton density at very 
high energies: it consider only the emission process, but it does not take into account the 
possible recombination of partons which occurs 
at such high density. Indeed, the BFKL is a linear equation.
The recombination phenomena of partons occurring at very high energies are governed by non-linear effects (coherency effects) 
which can be taken into account only by a non-linear equation. 
Furthermore, in order to recover unitarization of the theory, the system eventually evolves towards a saturation region\cite{saturation}.
Since coherency effects at high energies are important, the Breit frame introduced above is not anymore
a suitable frame in describing DIS processes. We consider the so called \textit{dipole frame}: we perform a Lorentz boost
in a frame in which the hadron still carries most of the energy, but the virtual photon has enough energy to split into a quark anti-quark
($q\bar{q}$) pair in the color singlet state long before the scattering. In this frame the lifetime of the $q\bar{q}$ 
fluctuation is much larger then the interaction time between this pair and the hadron. 
The \textit{dipole frame} is a natural frame work for the description of multiple scatterings which as we have
seen become important at high energy. We will describe the propagation of a $q\bar{q}$ pair 
at high energy through a medium generated by color field and we will show that
a semi-classical approach to the description of the DIS at small-$x_B$ leads to
a non-linear equation: the Balitsky-Kovchegov (BK) equation\cite{ba96,yura}.

\section{High-energy operator product expansion}
Before turning into the discussion of the semi-classical approach to DIS, let us remind ourselves some usual techniques 
used in the description of DIS in
the Bjorken regime. This will allow us to introduce similar technique for the description of DIS in the Regge limit.
The Wilson Operator Product Expansion\cite{Wilson:1969zs} (OPE) is a powerful way to study the ${\rm T}$ product of two electromagnetic currents 
${\rm T}\{j^\mu(x)j^\nu(y)\}$ at light-like separation $x^2\rightarrow 0$. The expansion is in terms of coefficient functions, 
pertubatively calculable, and matrix element of operators composed by fundamental fields such as quarks and 
gluons, not perturbatively calculable. An alternative way to the usual OPE is represented by the expansion of the 
${\rm T}$ product of two electromagnetic currents in terms of non-local string operators\cite{Balitsky:1983sw}.
This expansion allows one to extract in a more efficient way higher twist corrections to DIS.
As in the case of the usual Wilson expansion, all the singularities are included in the coefficients in front of such non-local operators;
the expansion in power of $(x-y)^\mu$ of the non local string operator gives back the Wilson OPE in local operators, and
in order to separate the higher twist effects one has to expand such non-local string operators on the light-cone.
Matrix elements of such string operators between two hadron state give the structure function of a given state\cite{Efremov:1979qk} and their
evolution\cite{Balitsky:1987bk} is governed by DGLAP type equation.
Furthermore, since this technique is in coordinate space it preserves explicitly gauge, Lorentz and conformal invariance.

\begin{figure}
\begin{center}
\hspace{-0.7cm}
\includegraphics[width=30mm]{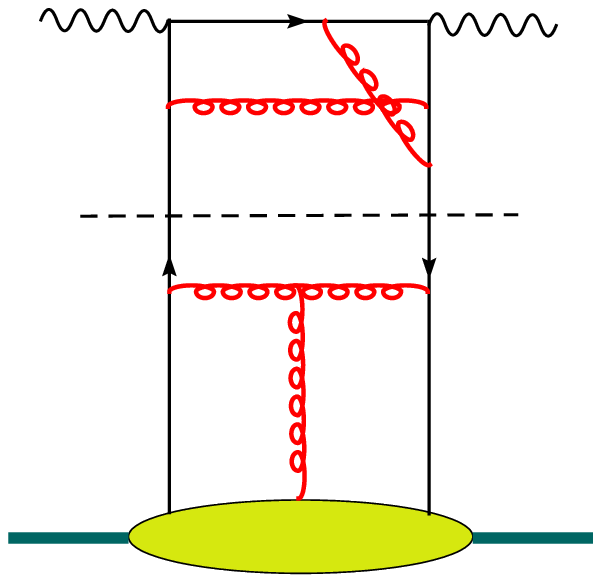}
\hspace{0.7cm}
\includegraphics[width=12mm]{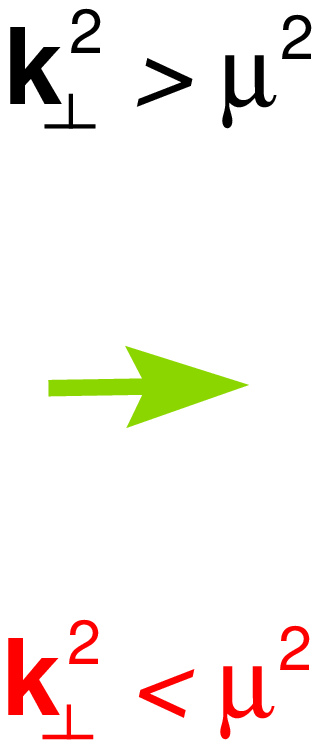}
\hspace{0.7cm}
\includegraphics[width=65mm]{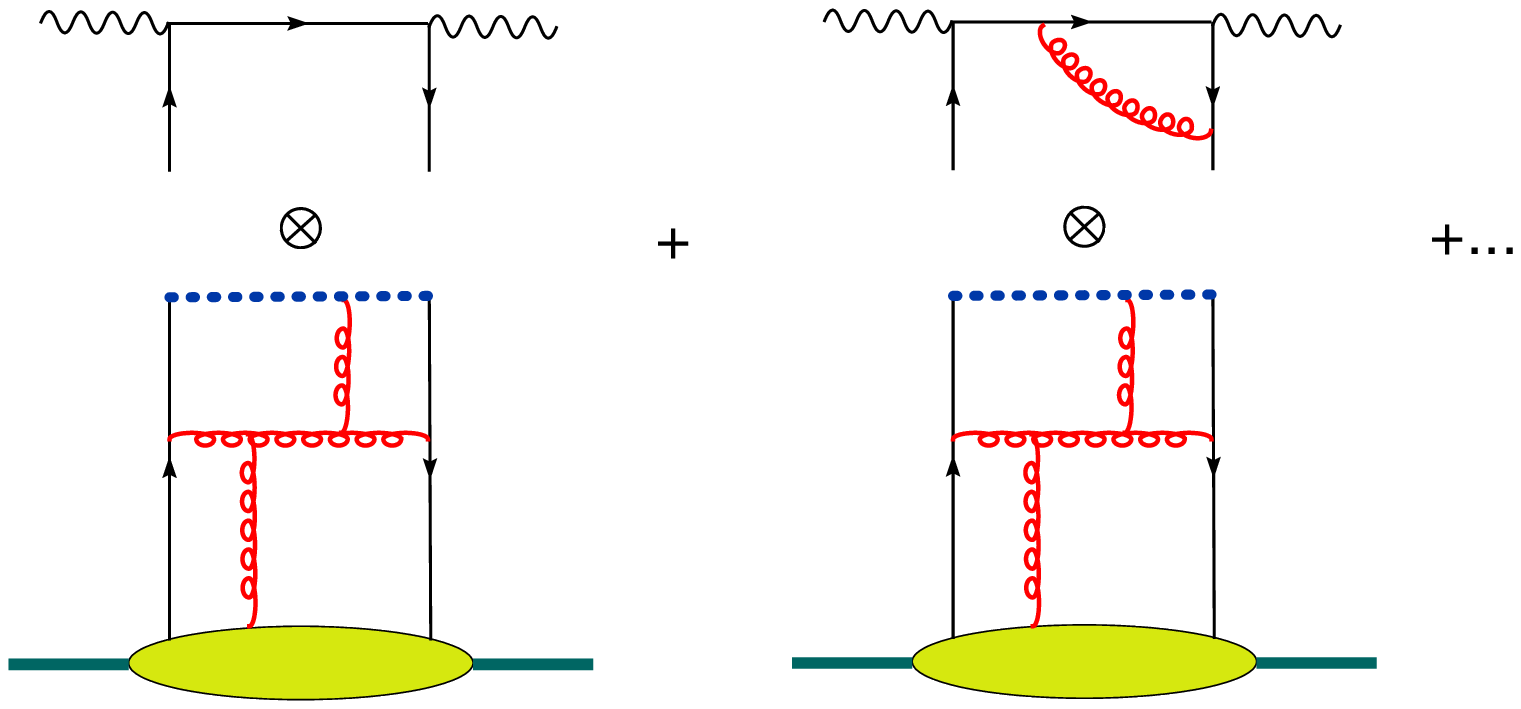}
\end{center}
\label{string-expan}
\caption{Expansion of the ${\rm T}$ product of two electromagnetic currents in terms of non-local string operators}
\end{figure}

As already mentioned above, due to the two different scales of transverse momentum $k_\perp$, 
in DIS it is natural to factorize the amplitude of the process
in the product of contributions of hard and soft parts coming from the
regions of small and large transverse momenta, respectively. Usually one
introduces a factorization scale $\mu$ to separate the hard and perturbative part from the soft and not perturbative part. 
The integrals over $k^2_\perp > \mu^2$
give the coefficient functions in front of light-cone operators while the contributions
from $k^2_\perp<\mu^2$ give matrix elements of these operators normalized at
the normalization point $\mu$. The evolution of such matrix elements with respect to the renormalization poit $\mu$ is the DGLAP 
evolution equation.
The expansion in terms of string operators is diagrammatically given in Fig. \ref{string-expan} 
where dotted lines represent the gauge link 
\begin{equation}
{\rm Pexp}\Big(ig\int_0^1 du(x-y)_\mu A^\mu(xu+(1-u)y)\Big)
\end{equation}

\begin{figure}[htb]
\psfig{file=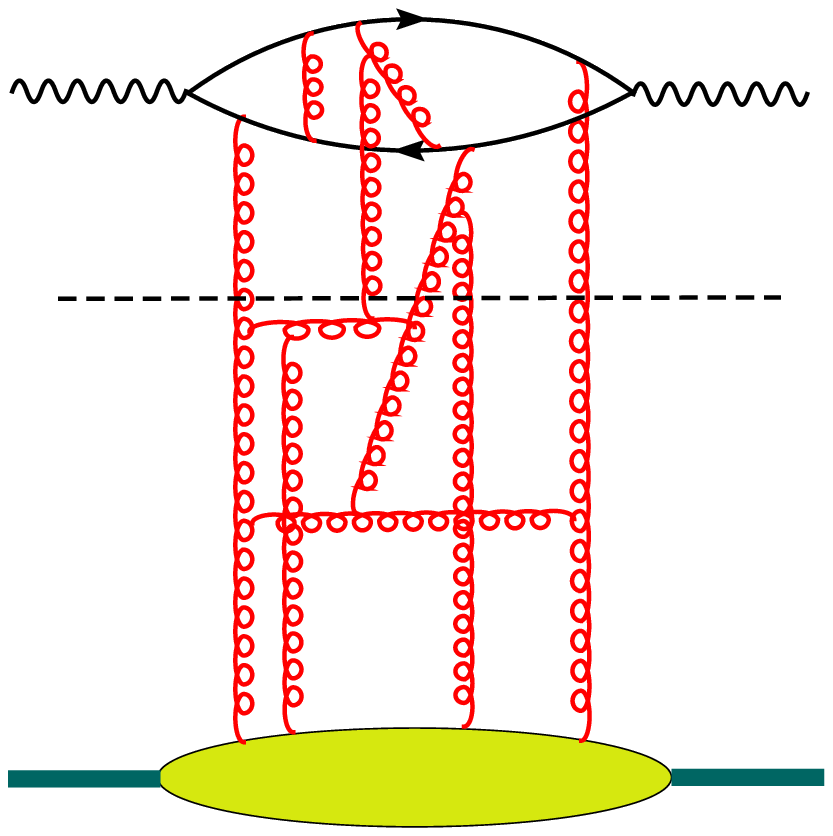,width=36mm} 
\hspace{2mm} 
\psfig{file=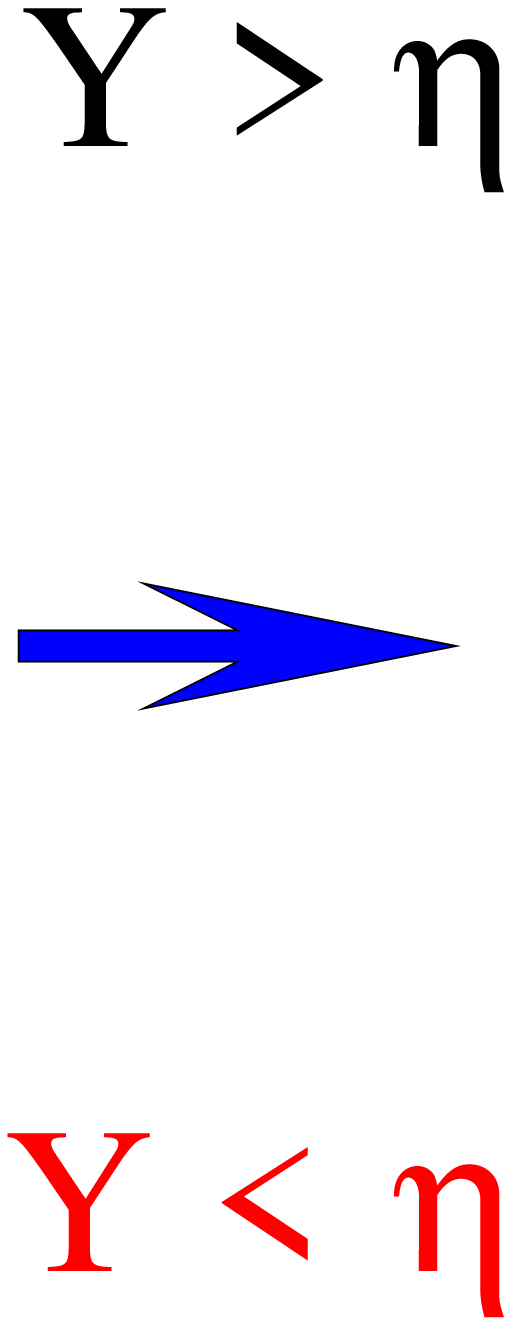,width=6mm} 
\hspace{2mm}
\psfig{file=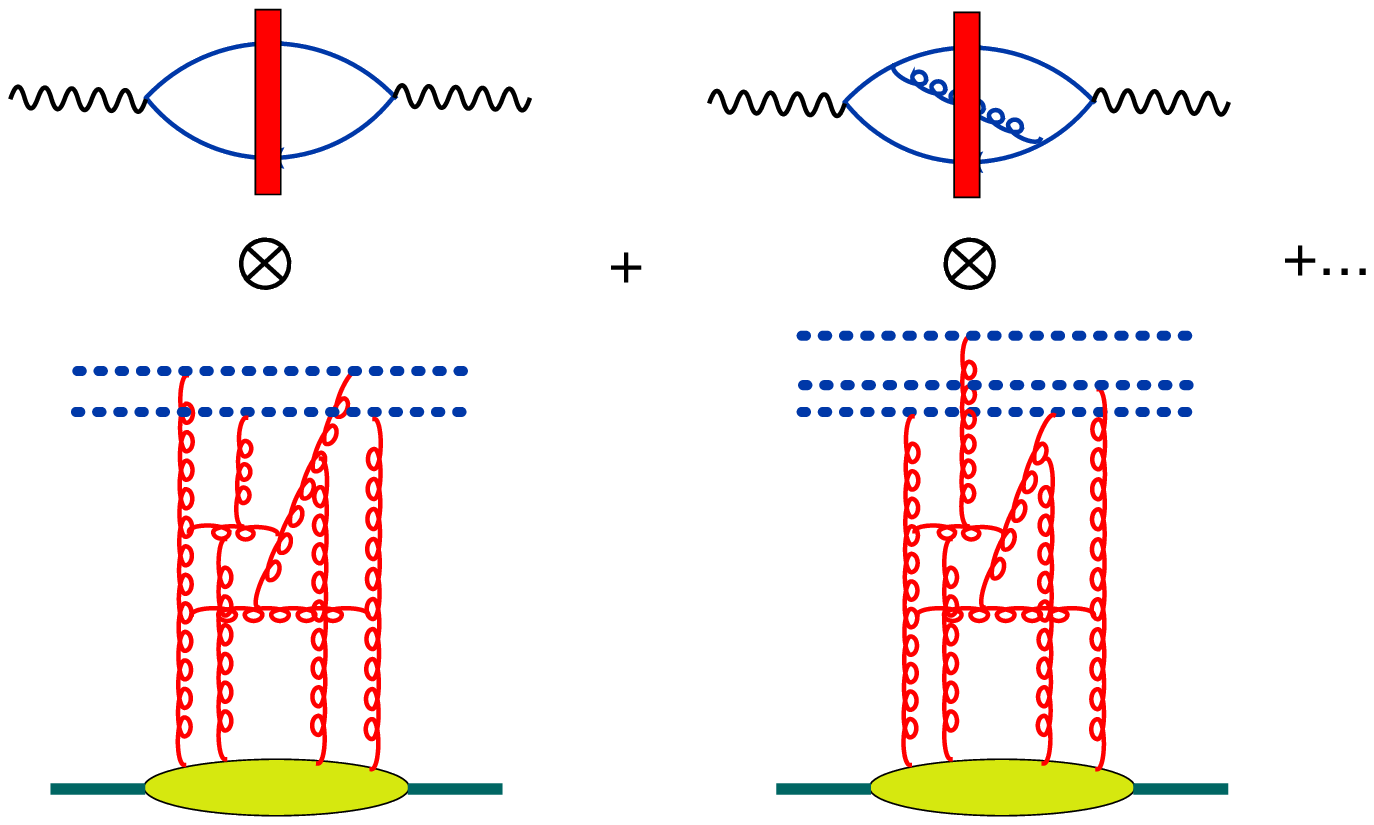,width=62mm}
\label{he-ex}
\caption{High-energy operator expansion in Wilson lines}
\end{figure}

Let us now turn our discussion to the case of high-energy (Regge) limit where all the transverse momenta
are of the same order of magnitude and therefore it is natural to introduce a different factorization scale
that is the rapidity. Factorization in
rapidity means that at high-energy scattering one introduces a rapidity divide $\eta$ which separate "fast" field from "slow" fields. Thus,
the amplitude of the process can be represented
as a convolution of contributions coming from fields with rapidity $\eta<Y$ ("fast" field) and contributions coming from fields
with rapidity $\eta>Y$ ("slow" fields). As in the case of the usual OPE, the integration over the field with rapidity $\eta<Y$ 
gives us the coefficients function while the integrations over the field with rapidity $\eta>Y$ are the matrix elements of the operators.
A general feature of high-energy scattering is that a fast particle moves along 
its straight-line classical trajectory and the only quantum effect is the eikonal phase 
factor acquired along this propagation path. In QCD, for the fast quark or gluon scattering off some target, 
this eikonal phase factor is a Wilson line - the infinite gauge link ordered along 
the straight line collinear to particle's velocity $n^\mu$:
\begin{equation}
U^\eta(x_\perp)={\rm Pexp}\Big\{ig\int_{-\infty}^\infty\!\! du ~n_\mu 
~A^\mu(un+x_\perp)\Big\},~~~~
\label{defU}
\end{equation}
Here $A_\mu$ is the gluon field of the target, $x_\perp$ is the transverse
position of the particle which remains unchanged throughout the collision, and the 
index $\eta$ labels the rapidity of the particle. Repeating the above argument for the target (moving fast in the spectator's frame) we see that 
particles with very different rapidity perceive each other as Wilson lines and
therefore these Wilson-line operators form
the convenient effective degrees of freedom in high-energy QCD (for a review, see Ref. \cite{mobzor}). 
The expansion of the ${\rm T}$ product of two electromagnetic currents at high-energy (Regge limit) is then in terms of Wilson lines 
\begin{eqnarray*}
&&\hspace{-8mm}
T\{\hat{j}_\mu(x)\hat{j}_\nu(y)\}=\int\! d^2z_1d^2z_2~I^{\rm LO}_{\mu\nu}(z_1,z_2)
{\rm Tr}\{\hat{U}^\eta_{z_1}\hat{U}^{\dagger\eta}_{z_2}\}
\nonumber\\
&&\hspace{-8mm}
+\int\! d^2z_1d^2z_2d^2z_3~I^{\rm NLO}_{\mu\nu}(z_1,z_2,z_3)
[ {\rm tr}\{\hat{U}^\eta_{z_1}\hat{U}^{\dagger\eta}_{z_3}\}{\rm tr}\{\hat{U}^\eta_{z_3}\hat{U}^{\dagger\eta}_{z_2}\}
-N_c{\rm tr}\{\hat{U}^\eta_{z_1}\hat{U}^{\dagger\eta}_{z_2}\}]
\label{high-energy-ex}
\end{eqnarray*}
We indicate operators by "hat". The diagrammatic version of eq. (\ref{high-energy-ex}) is given in Fig. \ref{he-ex}:
the dotted lines represent the Wilson lines while the coefficients functions are represented by the propagation of the $q\bar{q}$
pair in the background of shock-wave (represented in the figure by the red strip).

In Ref. \cite{nlobkN4} such expansion is performed in the ${\cal N}=4$ SYM theory for two BPS-protected currents ${\calo}\equiv{4\pi^2\sqrt{2}\over \sqrt{N_c^2-1}}
{\rm Tr\{Z^2\}}$ ($Z={1\over\sqrt{2}}(\phi_1+i\phi_2)$) and it is given by
\begin{eqnarray}
&&\hspace{-1mm}
(x-y)^4 T\{\halo(x)\halo(y)\}~
=~{(x-y)^4\over \pi^2(N_c^2-1)}\!\int\! d^2 z_{1\perp}d^2 z_{2\perp}
{(x_\ast y_\ast)^{-2}\over \calz_1^2\calz_2^2}~[{\rm Tr}\{\hat{U}^\eta_{z_1}\hat{U}^{\dagger\eta}_{z_2}\}]^{\rm conf}
\label{opeconfa}\nonumber\\
&&\hspace{-1mm}
-~{\alpha_s(x-y)^4\over 2\pi^4(N_c^2-1)}\!\int\! d^2 z_{1\perp}d^2 z_{2\perp} d^2 z_3
{z_{12}^2(x_\ast y_\ast)^{-2}\over z_{13}^2z_{23}^2\calz_1^2\calz_2^2}
\Big(\ln{a\sigma^2 s^2z_{12}^2\over 16 z_{13}^2z_{23}^2}\Big[{(x-z_3)^2\over x_\ast}-{(y-z_3)^2\over y_\ast}\Big]^2
\nonumber\\
&&\hspace{4mm}
-i\pi+2C\Big)
[ {\rm Tr}\{T^n\hat{U}^\eta_{z_1}\hat{U}^{\dagger\eta}_{z_3}T^n\hat{U}^\eta_{z_3}\hat{U}^{\dagger\eta}_{z_2}\}
-N_c {\rm Tr}\{\hat{U}^\eta_{z_1}\hat{U}^{\dagger\eta}_{z_2}\}]
\end{eqnarray}
where
\begin{eqnarray}
&&\hspace{-15mm}
[{\rm Tr}\{\hat{U}^\eta_{z_1}\hat{U}^{\dagger\eta}_{z_2}\}\big]^{\rm conf}={\rm Tr}\{\hat{U}^\eta_{z_1}\hat{U}^{\dagger\eta}_{z_2}\}\nonumber\\
&&\hspace{-15mm}
+{\alpha_s\over 2\pi^2}\!\int\! d^2 z_3~{z_{12}^2\over z_{13}^2z_{23}^2}
[ {\rm Tr}\{T^n\hat{U}^\eta_{z_1}\hat{U}^{\dagger\eta}_{z_3}T^n\hat{U}^\eta_{z_3}\hat{U}^{\dagger\eta}_{z_2}\}
-N_c {\rm Tr}\{\hat{U}^\eta_{z_1}\hat{U}^{\dagger\eta}_{z_2}\}]
\ln {az_{12}^2\over z_{13}^2z_{23}^2}
\label{confodipole}
\end{eqnarray}
is the \textit{composite dipole} with the conformal longitudinal cutoff in the next-to-leading order. The appearance of 
the \textit{composite operators} is due to the loss of conformal invariance of the Wilson line operator in the NLO.
Indeed, the light-like Wilson lines $U(x_\perp)$ are formally M\"obius invariant and consequently 
the leading-order BK equation is also conformal invariant. At NLO the Wilson line operator are divergent and its regularization introduces a dependence on the rapidity and conformal 
symmetry is lost. In order to restore the conformal invariance we redefine the operator ${\rm Tr}\{\hat{U}^\eta_{z_1}\hat{U}^{\dagger\eta}_{z_2}\}$ 
by adding suitable \textit{conterterms}. This procedure of finding the dipole with conformally regularized rapidity divergence is analogous
to the construction of the composite renormalized local operator by adding the appropriate counterterms order by order in perturbation theory.
\section{Leading order evolution of color dipoles} 
\begin{figure}
\centering
\includegraphics[width=0.9\textwidth]{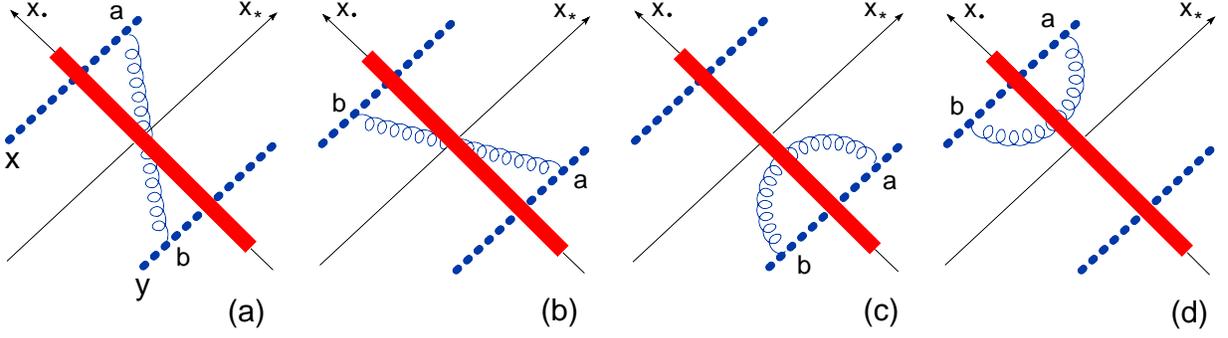}
\caption{Leading-order diagrams for the small-$x$ evolution of color dipole\label{bkevol}. Wilson lines are denoted by dotted lines.}
\end{figure}
Let us consider now the deep inelastic scattering from a hadron at small $x_B$.
As previously explained, the virtual photon decomposes into a pair of fast quarks 
moving along straight lines separated by some transverse distance.
The propagation of this quark-antiquark pair reduces to the 
``propagator of the color dipole'' that is two Wilson lines ordered along 
the direction collinear to quarks' velocity: $U(x_\perp)U^\dagger(y_\perp)$. 
The structure function of a hadron is proportional to a matrix element of this color dipole operator
\begin{equation}
\hat{\cal U}^\eta(x_\perp,y_\perp)=1-{1\over N_c}
{\rm Tr}\{\hat{U}^\eta(x_\perp)\hat{U}^{\dagger\eta}(y_\perp)\}
\label{fla1}
\end{equation}
switched between the target's states ($N_c=3$ for QCD). The gluon parton density is 
approximately
\begin{equation}
x_BG(x_B,\mu^2=Q^2)~
\simeq ~\left.\langle p|~\hat{\cal U}^\eta(x_\perp,0)|p\rangle
\right|_{x_\perp^2=Q^{-2}}
\label{fla2}
\end{equation}
where $\eta=\ln{1\over x_B}$.
The energy dependence of the structure function is translated then into 
the dependence of the color dipole on the slope of the Wilson lines determined by the rapidity $\eta$.
Thus, the small-x behavior of the structure functions is governed by the 
rapidity evolution of color dipoles\cite{dipole}. 
At relatively high energies and for sufficiently small dipoles we can use the leading logarithmic approximation (LLA)
where $\alpha_s\ll 1$, $\alpha_s\ln x_B\sim 1$ and get the non-linear BK evolution equation for the color
dipoles\cite{ba96,yura}:
\begin{eqnarray}
&&\hspace{-1.2cm}
{d\over d\eta}~\hat{\cal U}(x,y)=
{\alpha_sN_c\over 2\pi^2}\!\int\!d^2z~ {(x-y)^2\over(x-z)^2(z-y)^2}
[\hat{\cal U}(x,z)+\hat{\cal U}(y,z)\nonumber\\
&&\hspace{5cm}
-\hat{\cal U}(x,y)-\hat{\cal U}(x,z)\hat{\cal U}(z,y)]
\label{bk}
\end{eqnarray}
The first three terms correspond to the linear BFKL evolution\cite{bfkl} 
and describe the parton emission while the last term is responsible for the parton annihilation. 
For sufficiently high $x_B$ the parton emission balances the parton annihilation so the partons reach the state of saturation \cite{saturation} with
the characteristic transverse momentum $Q_s$ growing with energy $1/x_B$.
\section{NLO evolution of color dipole}
The NLO evolution of color dipole in QCD performed in Ref. \cite{nlobk} is not expected to be M\"obius invariant due to the conformal anomaly leading 
to dimensional transmutation and running coupling constant. However, the NLO BK equation in QCD has an additional 
term violating M\"obius invariance and not related to the conformal anomaly. 
To understand the relation between the high-energy behavior of amplitudes and M\"obius invariance of Wilson lines, 
it is instructive to consider the conformally invariant ${\cal N}=4$ super Yang-Mils theory. 
This theory was intensively studied in recent years due to the fact that at large coupling constants 
it is dual to the IIB string theory in the AdS$_5$ background. In the light-cone limit,
the contribution of scalar operators to Maldacena-Wilson line\cite{mwline} 
vanishes so one has the usual Wilson line constructed from gauge fields and therefore the LLA evolution
equation for color dipoles in the ${\cal N}=4$ SYM has the same form as (\ref{bk}).
At the NLO level, the contributions from gluino and scalar loops enter the picture.
As we mentioned above, formally the light-like Wilson lines are M\"obius invariant.
We also mentioned that the light-like Wilson lines are divergent in the longitudinal direction and moreover, it is exactly the evolution 
equation with respect to this longitudinal cutoff which governs the high-energy behavior of amplitudes. 
At present, it is not known how to find the conformally invariant cutoff in the longitudinal direction. 
When we use the non-invariant cutoff 
we expect, as usual, the invariance to hold in the leading order but to be
violated in higher orders in perturbation theory. In the calculation of Ref. \cite{nlobkN4} we restrict the longitudinal momentum of the gluons composing Wilson lines, 
and with this non-invariant cutoff the NLO evolution equation in QCD has extra non-conformal parts not related to the running of coupling constant.
Similarly, there will be non-conformal parts coming from the longitudinal cutoff of Wilson lines in the ${\cal N}=4$ SYM equation.
Here we present the result for the NLO evolution of the color dipole in ${\cal N}=4$ SYM ( performed in Ref. \cite{nlobkN4})
in the adjoint representation
(we use notations $z_{ij}\equiv z_i-z_j$ and $(T^a)_{bc}=-if^{abc}$)
\begin{eqnarray}
&&\hspace{-4mm}
{d\over d\eta}\big[{\rm Tr}\{\hat{U}^\eta_{z_1}\hat{U}^{\dagger\eta}_{z_2}\}\big]^{\rm conf}~
\label{nlobksymconf}\\
&&\hspace{-4mm}
={\alpha_s\over \pi^2}
\!\int\!d^2z_3~
{z_{12}^2\over z_{13}^2 z_{23}^2}\Big[1-
{\alpha_sN_c\over 4\pi}{\pi^2\over 3}\Big]
\big[{\rm Tr}\{T^a\hat{U}^\eta_{z_1}\hat{U}^{\dagger\eta}_{z_3}T^a\hat{U}_{z_3}\hat{U}^{\dagger\eta}_{z_2}\} 
-N_c {\rm Tr}\{\hat{U}^\eta_{z_1}\hat{U}^{\dagger\eta}_{z_2}\}\big]^{\rm conf}
\nonumber\\
&&\hspace{-2mm} 
-{\alpha_s^2\over 4\pi^4}
\int \!d^2 z_3 d^2 z_4 {z_{12}^2\over z_{13}^2z_{24}^2z_{34}^2}
\Big\{2\ln{z_{12}^2z_{34}^2\over z_{14}^2z_{23}^2}
+\Big[1+{z_{12}^2z_{34}^2\over z_{13}^2z_{24}^2-z_{14}^2z_{23}^2}\Big]\ln{z_{13}^2z_{24}^2\over z_{14}^2z_{23}^2}\Big\}
\nonumber\\ 
&&\hspace{-2mm}
\times {\rm Tr}\{[T^a,T^b]\hat{U}^\eta_{z_1}T^{a'}T^{b'}\hat{U}^{\dagger\eta}_{z_2}
+ T^bT^a\hat{U}^\eta_{z_1} [T^{b'},T^{a'}]\hat{U}^{\dagger\eta}_{z_2}\}
[(\hat{U}^\eta_{z_3})^{aa'}(\hat{U}^\eta_{z_4})^{bb'}-(z_4\rightarrow z_3)]
\nonumber
\end{eqnarray}
where $a$ is an arbitrary dimensional constant.
In fact, $a(\eta)=ae^\eta$ plays the same role for the rapidity evolution as $\mu^2$ for the usual 
DGLAP evolution: the derivative ${d\over da}$ gives the 
evolution equation (\ref{nlobksymconf}).
The kernel in the r.h.s. of Eq. (\ref{nlobksymconf}) is obviously M\"obius invariant since it depends on two four-point conformal 
ratios ${z_{13}^2z_{24}^2\over z_{14}^2z_{23}^2}$ and ${z_{12}^2z_{34}^2\over z_{13}^2 z_{24}^2}$. 
In \cite{nlobkN4} we also demonstrate that Eq. (\ref{nlobksymconf}) 
agrees with forward NLO BFKL calculation of Ref. \cite{lipkot}.
Let us now present the calculation for the NLO BK kernel in the case of QCD
\begin{eqnarray}
&&\hspace{-5mm}
{d\over d\eta}\big[{\rm tr}\{ \hat{U}^\eta_{z_1} \hat{U}^{\dagger\eta}_{z_2}\}\big]^{\rm conf}~
=~{\alpha_s\over 2\pi^2}
\!\int\!d^2z_3~
{z_{12}^2\over z_{13}^2z_{23}^2}\Big[1
\nonumber\\ 
&&\hspace{-2mm}
+~{\alpha_s\over 4\pi}\Big[b(\ln {z_{12}^2\mu^2\over 4}+2C)
-b{z_{13}^2-z_{23}^2\over z_{12}^2}\ln{z_{13}^2\over z_{23}^2}+
\big({67 \over 9}-{\pi^2\over 3}\big)N_c-{10\over 9}n_f\Big]\nonumber\\
&&\hspace{-2mm} \times
\big[{\rm tr}\{\hat{U}^\eta_{z_1}\hat{U}^{\dagger\eta}_{z_3}\}{\rm tr}\{\hat{U}^\eta_{z_3}\hat{U}^{\dagger\eta}_{z_2}\}
-N_c {\rm tr}\{\hat{U}^\eta_{z_1}\hat{U}^{\dagger\eta}_{z_2}\}\big]^{\rm conf}
\nonumber\\
&&\hspace{-2mm} 
+~{\alpha_s^2\over 16\pi^4}
\int \!{d^2 z_3d^2 z_4\over z_{34}^4}\Bigg[ \Big\{\Big(
-2+2{z_{12}^2z_{34}^2\over z_{13}^2z_{24}^2}\ln{z_{12}^2z_{34}^2\over z_{14}^2z_{23}^2}\nonumber\\
&&\hspace{-2mm} 
+
\Big[{z_{12}^2z_{34}^2\over z_{13}^2z_{24}^2}
\big(1+{z_{12}^2z_{34}^2\over z_{13}^2z_{24}^2-z_{14}^2z_{23}^2}\big)
+{2z_{13}^2z_{24}^2-4z_{12}^2z_{34}^2\over z_{13}^2z_{24}^2-z_{14}^2z_{23}^2}
\Big]\ln{z_{13}^2z_{24}^2\over z_{14}^2z_{23}^2}\Big)
+~\big(z_3\leftrightarrow z_4\big)\Big\}
\nonumber\\
&&\hspace{-2mm} \times~\big[\big({\rm tr}\{\hat{U}^\eta_{z_1}\hat{U}^{\dagger\eta}_{z_3}\}{\rm tr}\{\hat{U}^\eta_{z_3}\hat{U}^{\dagger\eta}_{z_4}\}
{\rm tr}\{\hat{U}^\eta_{z_4}\hat{U}^{\dagger\eta}_{z_2}\}
-{\rm tr}\{\hat{U}^\eta_{z_1}\hat{U}^{\dagger\eta}_{z_3}\hat{U}^\eta_{z_4}
\hat{U}^{\dagger\eta}_{z_2}\hat{U}^\eta_{z_3}\hat{U}^{\dagger\eta}_{z_4}\}
\big)
-(z_4\rightarrow z_3)\big]
\nonumber\\
&&\hspace{-2mm} 
+~{z_{12}^2z_{34}^2\over z_{13}^2z_{24}^2}
\Big\{2\ln{z_{12}^2z_{34}^2\over z_{14}^2z_{23}^2}
+\Big[1+{z_{12}^2z_{34}^2\over z_{13}^2z_{24}^2-z_{14}^2z_{23}^2}\Big]\ln{z_{13}^2z_{24}^2\over z_{14}^2z_{23}^2}\Big\}\nonumber\\
&&\hspace{-2mm} \times
\big({\rm tr}\{\hat{U}^\eta_{z_1}\hat{U}^{\dagger\eta}_{z_3}\}{\rm tr}\{\hat{U}^\eta_{z_3}\hat{U}^{\dagger\eta}_{z_4}\}
{\rm tr}\{\hat{U}^\eta_{z_4}\hat{U}^{\dagger\eta}_{z_2}\}
-z_3\leftrightarrow z_4\big)\Bigg]
\nonumber\\
&&\hspace{-2mm} 
+~{\alpha^2_sn_f \over 2\pi^4}\!\int\!{d^2z_3 d^2z_4\over z_{34}^4}
\Big\{2-{z_{13}^2z_{24}^2+z_{23}^2z_{14}^2- z_{12}^2z_{34}^2\over z_{13}^2z_{24}^2-z_{14}^2z_{23}^2}
\ln{z_{13}^2 z_{24}^2\over z_{14}^2z_{23}^2}\Big\}\nonumber\\
&&\hspace{-2mm} \times
{\rm tr}\{t^a\hat{U}^\eta_{z_1}t^b\hat{U}^{\dagger\eta}_{z_2}\}
{\rm tr}\{t^a\hat{U}^\eta_{z_3}t^b(\hat{U}^{\dagger\eta}_{z_4}-\hat{U}^\eta_{z_3})\}
\nonumber\\
&&\hspace{26mm}
\label{nlobksymqcd}
\end{eqnarray}
where $b={11\over 3}N_c-{2\over 3}n_f$. As previously advertised the NLO BK kernel in QCD for the composite operators
resolves in a sum of the conformal part and the running-coupling part.
\section{Conclusions}
We have discussed the DIS scattering in the Bjorken Limit and in the Regge limit; 
we have briefly reviewed the standard techniques used to study DIS, that is OPE in local operators and in non-local string operator. This
allowed us to introduce the OPE at high-energy of the ${\rm T}$ of two electromagnetic current in terms of Wilson lines operators. 
We have observed the BFKL equation gives us a nice prediction of the increase of the parton density at high energy, but it fails 
at very high energy since it would violate the unitarity condition.
The recombination phenomena of partons occurring at very high energies are governed by non-linear effects (coherency effects) 
which can be taken into account only by a non-linear equation. 
Furthermore, we have seen that in order to recover the unitarization of the theory, the system eventually evolves towards a saturation region.
Then Breit frame introduced above is not anymore
a suitable frame to describe DIS processes. Instead, we considered the so called \textit{dipole frame} 
which is a natural frame work for the description of multiple scattering which are relevant at high energy. 
The evolution of Wilson line operator ${\cal U}(x_\perp, y_\perp)$ is the needed non linear equation for the 
description of the coherency effect appearing at this regime. This evolution equation is the BK equation.
At the end we have presented the result for the NLO kernel of the BK equation in the case of QCD and in the ${N}=4$ SYM theory.
In order to restore conformal symmetry we have introduced the conformal composite operator.

\section*{Acknowledgments}
The author thanks the organizers, and in particular Prof. Radyushkin
for the warm hospitality and support received during the workshop.

\end{document}